\documentclass[12pt]{article}
\title{Singularity of Input-Output Matrices: Structural Linkages and Spectral Properties}
\author{Mohit Arora\thanks{Corresponding Author: Department of Economics, Union College, Schenectady, NY 12308. Email: marorajnu@gmail.com.} \hspace{3cm} Deepankar Basu\thanks{Department of Economics, University of Massachusetts Amherst, 310 Crotty Hall, 412 N. Pleasant Street, Amherst MA 01002. Email: dbasu@econs.umass.edu.}}
\date{July 27, 2026}

\usepackage{amsmath,amsthm,amssymb}
\usepackage[margin=1in]{geometry}
\usepackage{titlesec}
\titleformat{\section}{\normalfont\normalsize\bfseries\filcenter}{\thesection.}{0.5em}{}
\titleformat{\subsection}{\normalfont\normalsize\bfseries\filcenter}{\thesubsection.}{0.5em}{}
\usepackage[round]{natbib}
\usepackage{setspace}
\usepackage{changepage}
\usepackage{rotating}
\usepackage{booktabs,caption}
\usepackage[flushleft]{threeparttable}
\usepackage{url}

\newtheorem{theorem}{Theorem}

\begin{document}
	\maketitle
	\begin{abstract}
    Analyzing $5$ historical and $670$ contemporary input-output (IO) tables, we show that all of these IO matrices are singular. We map this phenomenon onto underlying production network topologies. While singularity trivially emerges from sectors lacking intermediate forward or backward linkages, it could frequently persist due to deeper economic properties: proportional cost structures across sectors, isolated economic sub-networks, and supply-chain hierarchies lacking intra-industry transactions. As a subset of these matrices is non-diagonalizable, we leverage this singularity to derive simple, rank and eigenvalue based diagnostic criteria for diagonalizability. This provides an efficient computational test for network shock propagation models using eigendecomposition.
	\end{abstract}
		
	\newpage
	\section{Introduction}
    In this paper, we present a fundamental empirical feature of input-output (IO) tables: the vast majority of IO matrices are singular.\footnote{Throughout this paper, the IO matrix refers to the expenditure share matrix defined in Section \ref{sec:io-sing}.} We show that singularity is universal across both historical and contemporary IO matrices. This feature is often observed by IO matrices having a full row or column of zeros or both. However, this is also evident in cases where a row or column is linearly dependent on other rows/columns.
    
    We explore the economic meaning of singularity in IO matrices, demonstrating how linear dependence maps directly onto underlying economic structures. Singularity trivially emerges when sectors lack intermediate forward or backward linkages, generating zero rows or columns. However, because zero vectors are a sufficient rather than necessary condition, singularity frequently persists in dense matrices too (where no all-zero rows or columns exist). In these cases, linear dependence signals deeper economic properties: proportional cost structures across sectors, isolated economic blocks trading exclusively in closed sub-networks, or supply-chain hierarchies with sectors that report zero intra-industry transactions.

	Recent literature has utilized eigendecomposition of the IO matrix to study the dynamic propagation of shocks through production networks \citep{liu2020dynamical,liu-2023-io}. Eigendecomposition relies on the requirement that the matrix of eigenvectors associated with the IO matrix is invertible—that is, the matrix must be \textit{diagonalizable}. Matrices that cannot be diagonalized are known as \textit{defective}.
	
	While standard linear algebra texts establish that distinct eigenvalues ensure diagonalizability \citep[chapter~5]{strang4th}, empirical IO matrices frequently feature repeated eigenvalues. Determining diagonalizability by computing eigenspace dimensions can be computationally cumbersome. We exploit the empirically observed pervasive singularity in IO matrices to provide simple, alternative diagnostic criteria for diagonalizability based on matrix rank and distinct non-zero eigenvalues.

    \section{Empirical Evidence on Singularity of IO Matrices}\label{sec:io-sing}
	
	For any economy, let $U$ denote the inter-industry transaction matrix and $<X>$ denote the diagonal matrix of industry-level gross outputs.\footnote{We drop industries with zero gross output to ensure that $<X>$ is invertible.} The IO expenditure share matrix is defined as:
	\begin{equation}\label{def:io-mat}
		A = U <X>^{-1}.
	\end{equation}
	
	We evaluate $675$ empirical IO matrices across multiple databases and historical periods to test for singularity ($\det(A) = 0$) and diagonalizability. While an IO matrix is trivially singular if it contains an all-zero row or column, this is merely a sufficient condition. To determine whether the pervasive singularity we observe is driven simply by these missing linkages or by deeper structural redundancy, we first evaluate the frequency of zero-vectors across our datasets.

	\subsection{Contemporary IO Tables: WIOD and U.S. BLS}
	\begin{enumerate}
		\item \textbf{WIOD (2016 Release):} Analyzing $56$-industry annual tables for $43$ countries from $2000$ to $2014$ ($645$ tables total) from the World Input-Output Database (WIOD), we find that \textit{every single IO matrix is singular}. %Testing for diagonalizability reveals that 638 out of $645$ tables satisfy the conditions for diagonalizability.
		\item \textbf{U.S. Economy (1997--2021):} Using highly disaggregated $192$-industry nominal tables from the U.S. Bureau of Labor Statistics (BLS), we find that the IO matrices are singular for \textit{all $25$ years}. %We find that the IO matrices are diagonalizable for $20$ out of the $25$ years.
	\end{enumerate}
	
	To unpack the drivers of this pervasive singularity, we analyze the frequency of zero-linkage vectors (zero rows representing a lack of intermediate forward linkages, and zero columns representing a lack of intermediate backward linkages). As reported in Table~\ref{table:freq_wiod}, $50.54\%$ of the WIOD matrices contain at least one zero row, and $65.89\%$ contain at least one zero column. Furthermore, when these zero vectors occur, they represent deeply entrenched structural features rather than temporary measurement gaps. Table~\ref{table:pers_wiod} demonstrates that the year-to-year persistence of the zero rows and columns in the WIOD dataset exceeds $99\%$. 
    
    Crucially, however, Table \ref{table:freq_wiod} also reveals that zero-linkages do not fully explain this pervasive singularity: A substantial number of WIOD matrices contain absolutely no zero rows or columns. The fact that these fully dense matrices remain singular indicates that deeper, structural redundancies are at play—an economic phenomenon we unpack in Section \ref{sec:econ-meaning}.

	\begin{table}[h]
		\centering
		\begin{threeparttable}
			\caption{Frequency of Zero-Entries in WIOD IO Tables}
			\label{table:freq_wiod}
			\begin{tabular}{lcc} 
				\\ [-1.8ex] \hline \\[-1.8ex]
				Metric & Count & Percentage \\ 
				\hline \\[-1.8ex]  
				Total IO Matrices Analyzed & $645$ & $100\%$ \\ 
				Matrices with at least one Zero Row & $326$ & $50.54\%$ \\ 
				Matrices with at least one Zero Column & $425$ & $65.89\%$ \\  
				Matrices where Zero Rows and Columns are exactly the same sectors & $303$ & $46.98\%$ \\ 
				\hline \\[-1.8ex] 
			\end{tabular}
		\end{threeparttable}
	\end{table}

	\begin{table}[h]
		\centering
		\begin{threeparttable}
			\caption{Sector-Level Persistence of Zero-Entries (WIOD)}
			\label{table:pers_wiod}
			\begin{tabular}{lc} 
				\\ [-1.8ex] \hline \\[-1.8ex]
				Structural Feature & Year-to-Year Persistence \\ 
				\hline \\[-1.8ex]  
				Zero Rows & $99.10\%$ \\ 
				Zero Columns & $99.25\%$ \\ 
				Zero Both (Symmetric Zeros) & $99.67\%$ \\ 
				\hline \\[-1.8ex] 
			\end{tabular}
		\end{threeparttable}
	\end{table}

	The U.S. BLS matrices exhibit a distinct network topology. As shown in Table~\ref{table:freq_usa}, $100\%$ of the highly disaggregated U.S. tables possess at least one zero row, yet none contain a single zero column. Consistent with the WIOD data, Table~\ref{table:pers_usa} confirms that these zero rows are highly persistent across the $25$-year sample. 

	\begin{table}[h]
		\centering
		\begin{threeparttable}
			\caption{Frequency of Zero-Entries in U.S. BLS IO Tables}
			\label{table:freq_usa}
			\begin{tabular}{lcc} 
				\\ [-1.8ex] \hline \\[-1.8ex]
				Metric & Count & Percentage \\ 
				\hline \\[-1.8ex]  
				Total IO Matrices Analyzed & $25$ & $100\%$ \\ 
				Matrices with at least one Zero Row & $25$ & $100\%$ \\ 
				Matrices with at least one Zero Column & $0$ & $0\%$ \\ 
				Matrices where Zero Rows and Cols are exactly the same sectors & $0$ & $0\%$ \\ 
				\hline \\[-1.8ex] 
			\end{tabular}
		\end{threeparttable}
	\end{table}

	\begin{table}[h]
		\centering
		\begin{threeparttable}
			\caption{Sector-Level Persistence of Zero-Entries (U.S. BLS)}
			\label{table:pers_usa}
			\begin{tabular}{lc} 
				\\ [-1.8ex] \hline \\[-1.8ex]
				Structural Feature & Year-to-Year Persistence \\ 
				\hline \\[-1.8ex]  
				Zero Rows & $100\%$ \\ 
				Zero Columns & N/A \\  
				Zero Both (Symmetric Zeros) & N/A \\ 
				\hline \\[-1.8ex] 
			\end{tabular}
		\end{threeparttable}
	\end{table}

	\subsection{Historical IO Tables}
	Table \ref{table:freq_hist} reports frequency of zero vectors in $5$ historical IO tables -- IO matrices for Japan ($1935$), Germany ($1936$), and three Indian IO tables from the $1950$s ($1951$--$52$, $1953$--$54$, and $1955$--$56$).\footnote{The input-output table for Japan in 1935 was made available by Liu and Tsyvinski (2020) in Appendix
C of their paper while we found the German 1936 input-output table from Fremdling and Staeglin (2014)
following Liu and Tsyvinski (2020)’s references. The Indian input-output tables for 1951-52, 1953-54 and
1955-56 were found in IARNIW (1960), UN (1961) and Chakraverti (1968) respectively. These input-output
tables are available upon request.} All of these historical matrices feature both zero rows and zero columns.\footnote{Persistence analysis is not done on the historical IO tables because of a lack of availability of continuous time series data.}

	\begin{table}[h]
		\centering
		\begin{threeparttable}
			\caption{Frequency of Zero-Entries in Historical IO Tables}
			\label{table:freq_hist}
			\begin{tabular}{lcc} 
				\\ [-1.8ex] \hline \\[-1.8ex]
				Metric & Count & Percentage \\ 
				\hline \\[-1.8ex]  
				Total IO Matrices Analyzed & $5$ & $100\%$ \\ 
				Matrices with at least one Zero Row & $5$ & $100\%$ \\ 
				Matrices with at least one Zero Column & $5$ & $100\%$ \\ 
				Matrices where Zero Rows and Cols are exactly the same sectors & $2$ & $40\%$ \\ 
				\hline \\[-1.8ex] 
			\end{tabular}
		\end{threeparttable}
	\end{table}

	\section{Economic Intuition: What Does Singularity Tell Us?}\label{sec:econ-meaning}
	
	The universal presence of singularity in empirical IO tables is not merely a mathematical curiosity; it contains direct economic intuition regarding the structure of production networks. Singularity of a square matrix ($\det(A) = 0$) is equivalent to linear dependence among the rows or columns of $A$. Economically, this can arise from several structural sources:

\subsection{Zero Linkages}
A column of zeros in the matrix $A$ indicates a sector that purchases zero intermediate inputs from domestic industries (lacking backward linkages). Conversely, a row of zeros represents a sector that sells zero intermediate output to other domestic industries (lacking forward linkages), supplying its goods or services exclusively to final demand. 

Depending on how these zero-linkages align, they create specific network topologies that reduce the matrix rank and force singularity of the matrix $A$:

\begin{itemize}
    \item \textbf{Enclave Sectors (Total Isolation):} When a sector exhibits both zero intermediate inputs and zero intermediate sales, it functions as an enclave, entirely disconnected from the domestic inter-industry supply chain. For example, in the WIOD database, ``Activities of households as employers; undifferentiated goods- and services-producing activities of households for own use'' frequently emerges as an enclave sector for certain countries and years.
    \item \textbf{Sinks (Zero Forward, Active Backward):} These sectors purchase intermediate inputs from the domestic economy but do not sell any output back into the intermediate network. For example, ``Federal electric utilities'' in the U.S. BLS tables function as terminal nodes that siphon inputs but route their entire output to final demand.
    \item \textbf{Sources (Active Forward, Zero Backward):} Conversely, some sectors supply intermediate goods to other industries but purchase zero intermediate inputs themselves, relying exclusively on primary inputs like labor or imported capital. For example, in the WIOD database, ``Activities auxiliary to financial services and insurance activities'' emerges as a source sector for Malta between 2000 and 2003.
\end{itemize}

\subsection{Structural Redundancy}
Even when every sector actively trades (i.e., the matrix $A$ lacks zero-linkage vectors as is the case with a substantial number of IO matrices in Table \ref{table:freq_wiod}), singularity can still arise due to deeper structural features of the economy. These redundancies are typically driven by the following economic phenomena:

\begin{itemize}
    \item \textbf{Proportional Cost Structures:} If two sectors have identical supply chain structures but differ only in scale (e.g., they buy from the exact same suppliers in the exact same proportions), their columns in the IO matrix $A$ will be linearly dependent multiples of one another \citep{hatanaka1952note}.
    \item \textbf{Hierarchical Supply Chains:} Perfectly one-way supply chains—where raw materials flow to intermediate goods, and then to final goods with no feedback loops—result in triangular matrices. Because the determinant of a triangular matrix is the product of its principal diagonal entries, such a supply chain is singular whenever at least one sector does not use its own output (yielding a zero on the principal diagonal).\footnote{There is a detailed discussion of hierarchical production networks in \cite{liu2019industrial}.}
    \item \textbf{Isolated Economic Blocks:} If the economy fractures into disconnected sub-economies that trade exclusively internally, the matrix takes on a block-diagonal structure \citep{miller2009input}. While isolation alone does not guarantee singularity, it restricts the rank space; if any of these closed-loop blocks contain duplicate processes or internal linear dependencies, the entire economy's IO matrix becomes singular.
\end{itemize}

	\section{Diagnostic Criteria for Diagonalizability of Singular Matrices}\label{sec:main}
	
	Given that empirical IO matrices are routinely singular, researchers can bypass standard eigenspace calculations by utilizing simplified conditions connecting matrix rank and non-zero eigenvalues. The mathematical proofs of these conditions are provided in Appendix \ref{sec:appendix}.

	\begin{theorem}\label{thm:nec} (Necessary Condition for Diagonalizability).
		Let $A$ be an $n \times n$ matrix of real numbers. If $A$ is diagonalizable, then $\text{rank}(A)$ equals the number of non-zero eigenvalues of $A$ (taking account of the algebraic multiplicity of each of the nonzero eigenvalues).
	\end{theorem}

	\begin{theorem}\label{thm:suf} (Sufficient Condition for Singular Matrix Diagonalizability).
		Let $A$ be an $n \times n$ matrix of real numbers with at least one zero eigenvalue. If
		\begin{enumerate}
			\item $\text{rank}(A) =$ number of non-zero eigenvalues, and 
			\item all non-zero eigenvalues are distinct,
		\end{enumerate}
		then $A$ is diagonalizable.
	\end{theorem}

	\noindent As demonstrated in Table~\ref{table:1}, these conditions instantly reveal that the 1950s Indian IO tables are non-diagonalizable because $\text{rank}(A) \neq \text{number of non-zero eigenvalues}$.

\begin{table}[h]
		\centering
		\begin{threeparttable}
			\caption{Structure and Diagonalizability of Historical IO Tables}
			\label{table:1}
			\begin{tabular}{lcccc} 
				\\ [-1.8ex] \hline \\[-1.8ex]
				Country-Year & Dimension ($n$) & Rank & Non-zero Eigenvalues & Status \\ 
				\hline \\[-1.8ex]  
				Japan 1935 & $23$ & $22$ & $22$ & Singular, Diagonalizable \\ 
				Germany 1936 & $40$ & $39$ & $39$ & Singular, Diagonalizable \\ 
				India 1951-52 & $36$ & $32$ & $31$ & Singular, Non-Diagonalizable \\ 
				India 1953-54 & $36$ & $32$ & $31$ & Singular, Non-Diagonalizable \\ 
				India 1955-56 & $36$ & $34$ & $33$ & Singular, Non-Diagonalizable \\ 
				\hline \\[-1.8ex] 
			\end{tabular}
		\end{threeparttable}
	\end{table}

Applying the same conditions to the contemporary WIOD and US IO tables, we find that 5 out of the 20 US IO tables are non-diagonalizable while 7 out of the 645 WIOD IO tables are non-diagonalizable.

    \section{Conclusion}\label{sec:conc}
This paper shows that singularity is an inherent feature of empirical input-output matrices reflecting the absence of linkages, proportional cost structures, isolated economic blocks, and supply-chain hierarchies lacking intra-industry transactions. Because these structural realities lead to rank-deficiency, the spectral properties of production networks cannot be taken for granted. For researchers attempting eigendecomposition to study dynamic network shock analysis, verifying diagonalizability is essential. The simplified rank-eigenvalue criteria presented in this note offer an efficient diagnostic tool for analyzing the diagonalizability of singular production networks.

	%\newpage

\appendix
	\section{Mathematical Proofs}\label{sec:appendix}

\begin{proof}[Proof of Theorem~\ref{thm:nec}]

According to the supposition of the theorem, the matrix $A$ is diagonalizable. 
Let $D$ be the diagonal matrix of eigenvalues of $A$, and let $P$ be the matrix of corresponding eigenvectors. 
Since $A$ is diagonalizable, the matrix of eigenvectors, $P$, is nonsingular. 
Hence, we can write: $D=P^{-1}AP$. 
Thus, we have $\operatorname{rank}(A)=\operatorname{rank}(P^{-1}A)=\operatorname{rank}(P^{-1}AP)=\operatorname{rank}(D)=$ number of nonzero eigenvalues (taking account of the algebraic multiplicity of each of the nonzero eigenvalues), where we have used the well known result that multiplication by nonsingular matrices does not change the rank of a matrix (Meyer, 2000, p. 210).
\end{proof}

\begin{proof}[Proof of Theorem~\ref{thm:suf}]
Suppose $A$ has $n-k$ nonzero eigenvalues and $0$ is repeated $k$ times as an eigenvalue of $A$, where $k\ge1$.

Let us first consider the zero eigenvalues.
Since $0$ is repeated $k$ times as an eigenvalue of $A$, its algebraic multiplicity $AM_{A}(0)=k$.
By the supposition of this theorem, $\operatorname{rank}(A)=n-k$. 
By the rank-nullity theorem (Strang, 2006, section 2.4), we have $\operatorname{rank}(A)+\operatorname{nullity}(A)=n$.
Thus, $\operatorname{nullity}(A)=n-\operatorname{rank}(A)=n-(n-k)=k$. 
Thus, the geometric multiplicity $GM_{A}(0)=\operatorname{nullity}(A)=k=AM_{A}(0)$.

Let us now consider the nonzero eigenvalues.
By the supposition of the theorem, each of the nonzero eigenvalues are distinct. 
Let $\lambda\ne0$ be an eigenvalue of $A$. 
By the supposition of the theorem, its algebraic multiplicity $AM_{A}(\lambda)=1$. Since the geometric multiplicity is weakly less than arithmetic multiplicity of any eigenvalue (Meyer, 2000, p. 512), we have $0<GM_{A}(\lambda)\le AM_{A}(\lambda)=1$.
This implies that $GM_{A}(\lambda)=AM_{A}(\lambda)=1$, because $GM_{A}(\lambda)$ is an integer.

Thus, we have proved that for each eigenvalue of $A$, whether it is zero or nonzero, geometric and algebraic multiplicity is equal.
This implies that $A$ is diagonalizable (Meyer, 2000, p. 512).
\end{proof}
\end{document}